\documentclass[fleqn,10pt]{wlscirep}
\usepackage[utf8]{inputenc}
\usepackage[T1]{fontenc}

\usepackage{amsmath,amssymb}
\usepackage{amsfonts}
\usepackage{algorithm}
\usepackage{algpseudocode}
\usepackage{setspace}

\newcommand{\Prob}[1]{\mathrm{Pr}\left(#1\right)}
\newcommand{\Ex}[1]{\mathbb{E}\left[#1\right]}
\newcommand{\regret}{{\rm regret}}
\newcommand{\ifunc}[1]{\mathbb{I}\left[#1\right]}
\newcommand{\integ}[4]{\int_{#1}^{#2} #3 \mathrm{d}#4}

\title{Arm order recognition in multi-armed bandit problem with laser chaos time series}

\author[1,*]{Naoki Narisawa}
\author[2]{Nicolas Chauvet}
\author[3]{Mikio Hasegawa}
\author[1,2,*]{Makoto Naruse}
\affil[1]{Department of Mathematical Engineering and Information Physics, Faculty of Engineering, The University of Tokyo, 7-3-1 Hongo, Bunkyo-ku, Tokyo 113-8656, Japan}
\affil[2]{Department of Information Physics and Computing, Graduate School of Information Science and Technology, The University of Tokyo, 7-3-1 Hongo, Bunkyo-ku, Tokyo 113-8656, Japan}
\affil[3]{Department of Electrical Engineering, Tokyo University of Science, 6-3-1 Niijuku, Katsushika-ku, Tokyo 125-0051, Japan}

\affil[*]{\url{narisawa-naoki682@g.ecc.u-tokyo.ac.jp}, \url{makoto_naruse@ipc.i.u-tokyo.ac.jp}}

\begin{abstract}
By exploiting ultrafast and irregular time series generated by lasers with delayed feedback, we have previously demonstrated a scalable algorithm to solve multi-armed bandit (MAB) problems utilizing the time-division multiplexing of laser chaos time series. Although the algorithm detects the arm with the highest reward expectation, the correct recognition of the order of arms in terms of reward expectations is not achievable. Here, we present an algorithm where the degree of exploration is adaptively controlled based on confidence intervals that represent the estimation accuracy of reward expectations. We have demonstrated numerically that our approach did improve arm order recognition accuracy significantly, along with reduced dependence on reward environments, and the total reward is almost maintained compared with conventional MAB methods. This study applies to sectors where the order information is critical, such as efficient allocation of resources in information and communications technology.

\end{abstract}

\begin{document}

\flushbottom
\maketitle
\thispagestyle{empty}
\doublespacing
\section*{Introduction} 

Chaos can be defined as random oscillations generated by deterministic dynamics \cite{lorenz1963deterministic}. Chaotic time series are very sensitive to initial conditions, which render long-term predictions unachievable unless infinite observation accuracy is attained in the beginning \cite{fischer2000synchronization}. The close relationship between lasers and chaos has been known for a long time; the output of a laser generates chaotic oscillations when a time-delayed optical feedback is injected back into the laser cavity \cite{uchida2012optical}. Laser chaos exhibits ultrafast dynamics beyond GHz regime/domain; hence, various engineering applications have been examined in the literature. Examples range from optical secure communication \cite{fischer2000synchronization} and fast physical random bit generation \cite{uchida2008fast} to secure key distribution using correlated randomness \cite{yoshimura2012secure}.

The present study relates to the application of laser chaos to a multi-armed bandit problem (MAB) \cite{naruse2018scalable}. Reinforcement learning (RL), a branch of machine learning along with supervised and unsupervised learning, studies optimal decision-making rules. It differs from other machine learning tasks (e.g. image recognition) as the notion of reward comes into play in RL. The goal of RL is to construct decision-making rules that maximize obtained rewards; hence, gaming AI is a well-known application of RL \cite{littman2015reinforcement}. In 2015, AlphaGo, developed by Google DeepMind, defeated a human professional Go player for the first time \cite{deepmind}. 

The MAB is a sequential decision problem of maximizing total rewards where there are $K~(>1)$ arms, or selections, whose reward probability is unknown. The MAB is one of the simplest problems in RL. In an MAB, a player can receive reward information that pertains only to the selected arm at each time step, so a player cannot obtain the reward information for a non-selected arm. The MAB exhibits a trade-off between exploration and exploitation. Sufficient exploration is necessary to estimate the best arm more accurately, but it accompanies low-reward arm selections. Hence, excessive exploration can lead to significant losses. Furthermore, to maximize rewards, one needs to choose the best arm (use the exploitation principle). However, if the search for the best arm fails, then a non-best option may be mistakenly chosen very likely. Therefore, it is important to balance exploration and exploitation.

An algorithm for the MAB using laser chaos time series has been proposed in 2018 \cite{naruse2018scalable}. This algorithm sets two goals: to maximize the total rewards and to identify the best arm. However, concerning real-world applications, maximizing the rewards and finding the optimal arm may not be enough to solve a problem. For example, there is a study to improve communication throughput by treating the channel selection in wireless communications as an MAB \cite{shungo2020dynamic}. Should we have multiple channel users, not all users can use the best channel simultaneously; accordingly, there may be situations where compromises must be made, i.e., other channels will be selected. Now it is obvious that particular channel performance ranking information would be useful when considering non-best channels.

Conversely, when there are no other users, a player (the single user) can simultaneously utilize top-ranking options to accelerate the communication ability, similar with the channel bonding in local area networks \cite{bejarano2013ieee}. The purpose of this study is to accurately recognize the order of the expected rewards of different arms using a chaotic laser time series and to minimize the reduction of accumulated rewards due to too detailed exploration.

\section*{Principles}
\subsection*{Definition and Assumption}
We consider an MAB problem in which a player selects one of $K$ slot machines, where $K = 2^M$ and $M$ is a natural number. The $K$ slot machines are distinguished by identities numbered from $0$ to $K-1$, which are also represented in $M$-bit binary code given by $S_1 S_2 \cdots S_M$ with $S_i\in\{0,1\}$ ($i = 1,...,M$). For example, when $K = 8~(\text{or }M = 3)$, the slot machines are numbered by $S_{1}S_{2}S_{3} = \{000,001,...,110,111\}$. In this study, we assume that $\mu_i\neq \mu_j$ if $i\neq j$, and we define the $k$-th max and $k$-th argmax operators as $\displaystyle\max^k\{\}$ and $\displaystyle\arg\max^k\{\}$. The variables used in the study are defined as described below:
\begin{itemize}
\item $X_{i}(n)$: Obtained reward from arm $i$ at time step $n$ (independent at each time step. $x_i(n)$ is observed value.)
\item $\mu_i:= \Ex{X_i(n)}$. (Consistent regardless of time step)
\item $\displaystyle \mu^*:=\max_{i}\mu_i, \quad i^*:= \arg\max_{i}\mu_i$
\item $T_i(n)$: Number of selections of arm $i$ by the end of time step $n$ ($t_i(n)$ is observed value).
\item $A(n)$: Arm selected at time step $n$ ($a(n)$ is the observed value).
\item $\displaystyle [k] := \arg\max_{i}^k \mu_i$: $k$-th best arm.
\end{itemize}

We estimate the arm order of reward expectations by calculating the sample mean of the accumulated reward at each time step. Specifically, the sample means of rewards obtained from arm $i$ by time step $n$ is calculated as follows:
\begin{align}
\hat{\mu}_i(n) = \frac{R_i(n)}{T_i(n)}, \quad \text{where}\quad
R_i(n) := \sum_{s=1}^nX_i(s)\cdot \ifunc{A(s)=i}
\end{align}
In each time step $n$, we estimated the arm $\displaystyle j:=\arg\max_i^k\hat{\mu}_i(n)$ as the $k$-th best arm. 

\subsection*{Time-division multiplexing of laser chaos}
The proposed method is based on the MAB algorithm reported in 2018 \cite{naruse2018scalable}. This method consists of the following steps: [STEP 1] decision making for each bit of the slot machines, [STEP 2] playing the selected slot machine, and [STEP 3] updating the threshold values.

\paragraph{[STEP 1] Decision for each bit of the slot machine}
First, the chaotic signal $s(t_1)$ measured at $t = t_1$ is compared to a threshold value denoted as $TH_1$. If $s(t_1) \ge TH_1$, then bit $S_1$ is assigned $1$. Otherwise, $S_1$ is assigned $0$. To determine the value of $S_k~(k = 2,...,M)$, the chaotic signal $s(t_k)$ measured at $t = t_k$ $(>t_{k-1})$ is compared to a threshold value denoted as $TH_{k,S_1\cdots S_{k-1}}$. If $s(t_{k}) \ge TH_{k,S_1\cdots S_{k-1}}$, then bit $S_k$ is assigned $1$. Otherwise, $S_k$ is assigned $0$. After this process, a slot machine with the number represented in a binary code $S_1\cdots S_M$ is selected.

\paragraph{[STEP 2] Slot machine play}
Play the selected slot machine.

\paragraph{[STEP 3] Threshold values adjustment}
If the selected slot machine yields a reward, then the threshold values are adjusted in a way that the same decision will be more likely to be selected. For example, if $S_1$ is assigned $0$ and the player gets a reward, then $TH_1$ should be increased because doing so increases the likelihood of getting $S_1 = 0$ again. All of the other threshold values involved in determining the decision (i.e. $TH_{2,S_1},...,TH_{M,S_1 \cdots S_{M-1}}$) are updated in the same manner.

If the selected slot machine does not yield a reward, then the threshold values are adjusted to make the same decision less likely to take place. For example, if $S_1$ is assigned $1$ and the player does not get a reward, then $TH_1$ should be increased because of the decreased likelihood of getting $S_1 = 1$. Again, all of the other threshold values involved in determining the decision (i.e. $TH_{2, S_1},...,TH_{M, S_1\cdots S_{M-1}}$) are updated in the same manner.

\subsection*{Arm order recognition algorithm with confidence intervals}
\begin{figure}[tbp]
\centering
\includegraphics[width=1.0\linewidth]{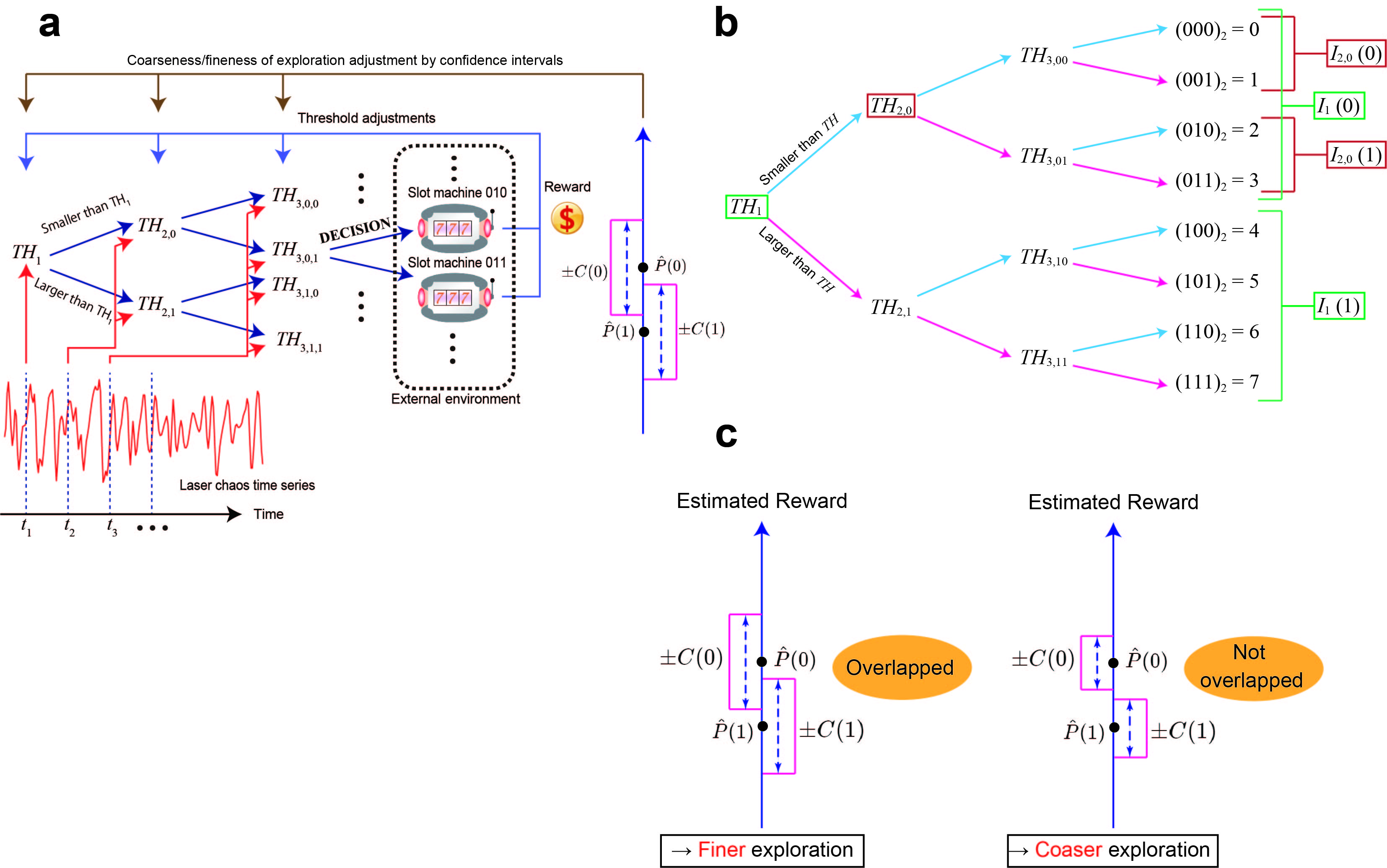}
\caption{Architecture of the proposed method with confidence intervals. \textbf{a} Solving the MAB with $K = 2^M$ arms using a pipelined arrangement of comparisons between thresholds and a series of chaotic signal sequences. {\bf{b}} Correspondence between threshold value $TH$ and a subset of arms $I(z)$ $(z\in\{0,1\})$ in the example of an eight-armed bandit problem. For each threshold $TH_*$, two types of arm set $I_*(0)$ and $I_*(1)$ are defined. {\bf c} Coarseness/fineness of exploration adjustment by confidence intervals. For each threshold $TH_*$, the fineness of the threshold adjustment is changed depending on whether two confidence intervals $\hat{P}_*(0;n)\pm C_*(0;n)$ and $\hat{P}_*(1;n)\pm C_*(1;n)$ are overlapped. A part of images in \textbf{a} is adapted from Naruse et al., Sci. Rep. 8, 10890 (2018). Copyright 2018 Author(s), licensed under a Creative Commons Attribution 4.0 License.}
\label{fig:ci_adjustment}
\end{figure}

\paragraph{Confidence intervals.}
For each threshold value $TH_{j, b_1\cdots b_{j-1}}$ ($j \in \{1, \cdots, M\}$, $b_1,...,b_{j-1}\in \{0, 1\}$) and $z\in \{0, 1\}$, the following values $\hat{P}(z;n)$ and $C(z;n)$ are calculated:
\begin{align}
&\hat{P}_{j, b_1\cdots b_{j-1}}(z; n) 
:= \frac{\sum_{i \in I_{j,b_1\cdots b_{j-1}}(z)}R_i(n)}{\sum_{i \in I_{j,b_1\cdots b_{j-1}}(z)}T_i(n)}, \quad
C_{j,b_1\cdots b_{j-1}}(z; n)
:= \gamma\cdot\sqrt{\frac{\log n}{\sum_{i\in I_{j,b_1\cdots b_{j-1}}(z)}T_i(n)}} \\
&I_{j,b_1\cdots b_{j-1}}(z) := \left\{\begin{array}{ll}
\{i~| ~\text{machine $i$ is available if } s(t_j) \ge TH_{j,b_1\cdots b_{j-1}}\} & (\text{if } z = 1) \\
\{i~| ~\text{machine $i$ is available if } s(t_j) < TH_{j,b_1\cdots b_{j-1}}\} & (\text{if } z = 0)
\end{array}\right.
\end{align}
$I_{j, b_1 \cdots b_{j-1}}(z)$ represents a subset of machine arms. 
If machine $i$ can be selected when the signal $s(t_j)$ is more than $TH_{j, b_1 \cdots b_{j-1}}$, then $i$ is included in $I_{j, b_1 \cdots b_{j-1}}(1)$. 
Otherwise, $i$ is not included in $I_{j, b_1 \cdots b_{j-1}}(1)$. 
In the same way, if machine $i$ can be selected when the signal $s(t_j)$ is less than or equal to $TH_{j, b_1 \cdots b_{j-1}}$, then $i$ is included in $I_{j, b_1 \cdots b_{j-1}}(0)$. 
Otherwise, $i$ is not included in $I_{j, b_1\cdots b_{j-1}}(0)$. 
For example, in the case of an eight-armed bandit problem (Fig. \ref{fig:ci_adjustment}b):
\begin{align*}
I_{1}(0) = \{0,1,2,3\},\quad
I_{2,0}(0) = \{0,1\}, \quad
I_{2,1}(1) = \{3,4\}, \quad
I_{3,00}(1) = \{1\}.
\end{align*}
$\hat{P}_{j, b_1\cdots b_{j-1}}(z; n)$ represents the sample means of rewards obtained from machines in $I_{j, b_1\cdots b_{j-1}}(z)$. 
$C_{j, b_1\cdots b_{j-1}}(z; n)$ represents the confidence interval width of the estimated value $\hat{P}_{j, b_1\cdots b_{j-1}}(z; n)$. 
The lower $C(z; n)$, the higher the estimation accuracy. 
Parameter $\gamma$ indicates the degree of exploration : a higher $\gamma$ means that more exploration is needed to reach a given confidence interval width.

\paragraph{Coarseness/fineness of exploration adjustments by confidence intervals.}
At each threshold $TH_{j, b_1\cdots b_{j-1}}$, if the two intervals 
\[
\left[\hat{P}(0; n)-C(0; n),~~
\hat{P}(0; n)+C(0; n)\right]
\quad \text{and} \quad
\left[\hat{P}(1; n)-C(1; n),~~
\hat{P}(1; n)+C(1; n)\right]
\]
are overlapped, we suppose there is a likelihood of a change in the order relationship between $\hat{P}(0; n)$ and $\hat{P}(1; n)$; that is, the order of $\hat{P}(0; n)$ and $\hat{P}(1; n)$ is not known yet. 
Therefore, the exploration process should be executed more carefully. 
Hence, the threshold value should be closer to $0$, which is a balanced situation, or we should perform further exploration, so that the threshold adjustment becomes finer. 
Conversely, if the two intervals are not overlapped, then we suppose a low likelihood of a wrong estimate of the order relationship between $\hat{P}(0; n)$ and $\hat{P}(1; n)$. Hence, we should continue exploration more coarsely so that the threshold adjustment will be accelerated. (Fig. \ref{fig:ci_adjustment}c)

\section*{Results}
\paragraph{Experimental settings.}
We have evaluated the performance of the methods for two cases: a four-armed bandit and an eight-armed bandit. 
First, the reward probability of each arm is assumed to follow the Bernoulli distribution: $\Prob{X_i=x} = \mu_i^x(1-\mu_i)^{1-x}$. 
Each reward environment $\nu:=(\mu_0,...,\mu_{K-1})$ is set to satisfy the following conditions: 
(i) $\forall i: \mu_i\in\{0.1,0.2,...,0.8.0.9\}$, 
(ii) $i\neq j \Rightarrow \mu_i\neq \mu_j$. 
In this experiment, a variety of assignments of reward probabilities $\nu$ satisfying the above conditions were prepared, and the performance was evaluated under every reward environment $\nu$. We have defined the reward, regret, and correct order rate (COR) as metrics to quantitatively evaluate the performance of the method.
\begin{align}
{\rm reward}(n) &:= \frac{1}{l_m}\sum_{l=1}^{l_m}\left[\sum_{s=1}^n x^{(l)}_{a^{(l)}(s)}(s)\right], \label{eq:def_reward}\\
{\rm regret}(n) &:= \frac{1}{l_m}\sum_{l=1}^{l_m}\left[\sum_{i\neq i^*} (\mu^*-\mu_i) t^{(l)}_i(n)\right], \label{eq:def_regret}\\
{\rm COR}(n) &:= \frac{1}{l_m}\sum_{l=1}^{l_m}\ifunc{\bigcap_{k=1,...,4}\left\{\arg\max_{i=1,...,K}^k\hat{\mu}_i^{(l)}(n) = [k]\right\}} \label{eq:def_cor}
\end{align}
where $n$ denotes number of time steps, $t_i(n)$ is the number of selections of arm $i$ up to time step $n$, and $l_m$ represents the number of measurements in one reward environment $\nu$. 
For the accuracy of arm order recognition, we considered the estimation accuracy of the top four arms regardless of the total number of arms. 
We prepared all 144 reward environments $\nu$ (all combinations satisfying the above conditions and $\max_{i\neq j}|\mu_i-\mu_j| = 0.3$) for the four-armed bandit problems and 100 randomly selected reward environments for the eight-armed bandit problems. 
The performances of four methods were compared: 
\textbf{RoundRobin} (all arms are selected in order at each time step), 
\textbf{UCB1} (method for maximizing the total rewards proposed in 2002 \cite{auer2002finite}), 
\textbf{Chaos} (previous method using the laser chaos time series \cite{naruse2018scalable}, only finding the best arm, not recognizing the order), and 
\textbf{Chaos-CI} (proposed method using laser chaos time series and with confidence intervals).

\begin{figure}[tbp]
\centering
\includegraphics[width=1.0\linewidth]{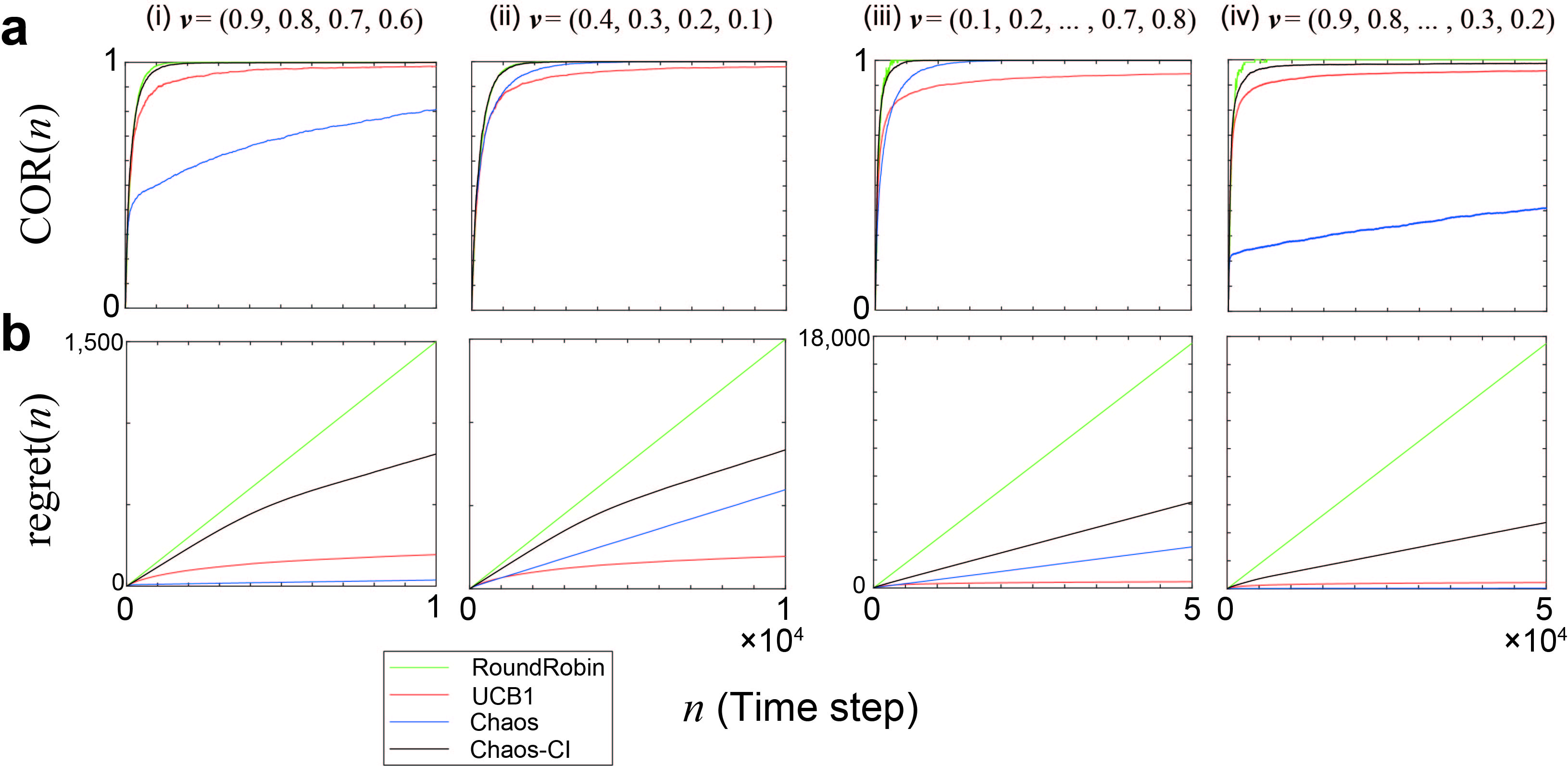}
\caption{Time evolution of metrics $\mathrm{COR}(n)$ and $\regret(n)$ under one reward environment $\nu$. Each column shows the results under each environment: the left two columns are in the four-armed bandit and the right two columns are in the eight-armed bandit. \textbf{a} $\mathrm{COR}(n)$ calculated as Eq. (\ref{eq:def_cor}). \textbf{b} $\regret(n)$ calculated as Eq. (\ref{eq:def_regret}).}
\label{fig:micro}
\end{figure}

\paragraph{Evaluation under one reward environment $\nu$.}
The curves in Figs. \ref{fig:micro}a and b show the time evolutions of $\regret(n)$ and $\mathrm{COR}(n)$, respectively, over $l_m = 12,000$ measurements under specific reward environments $\nu =  (\mu_0,...,\mu_{K-1})$. Specifically, columns (i) and (ii) pertain to the four-armed bandit problems defined by $\nu = (0.9,0.8,0.7,0.6)$ and $\nu = (0.4,0.3,0.2,0.1)$, whereas columns (iii) and (iv) depict the eight-armed bandit problems given by $\nu = (0.1,0.2,0.3,0.4,0.5,0.6,0.7,0.8)$ and $\nu = (0.9,0.8,0.7,0.6,0.5,0.4,0.3,0.2)$. The curves were colour coded for an easy method comparison. In the arm order recognition, Chaos-CI and RoundRobin presented high accuracy in the early time step. In terms of total reward, Chaos and UCB1 achieved the greatest rewards. 

\begin{figure}[tbp]
\centering
\includegraphics[width=1.0\linewidth]{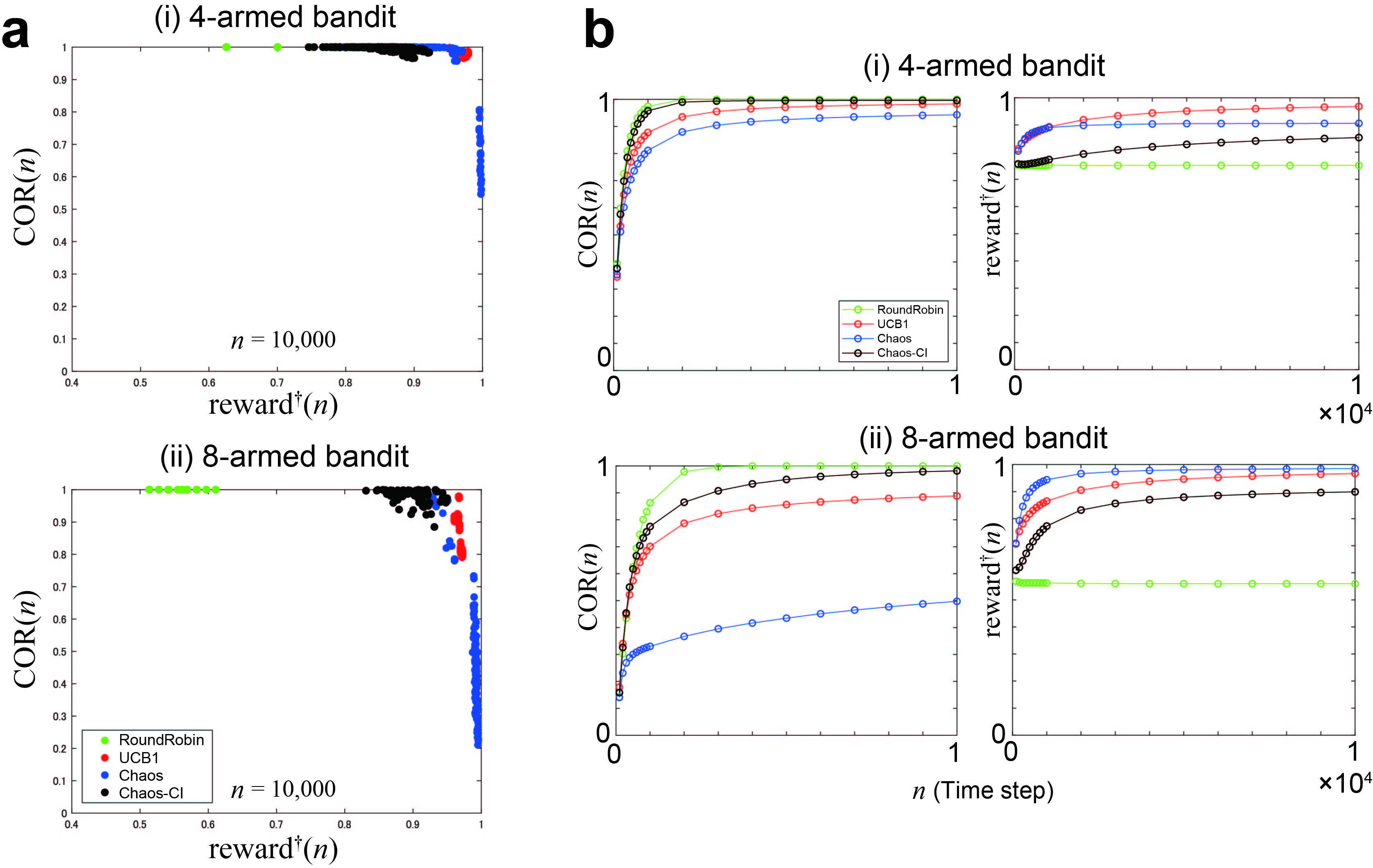}
\caption{Metrics over the whole reward environments prepared. \textbf{a} Each scatter plot represents a normalized reward $\mathrm{reward}^\dagger(n)$ and correct order rate $\mathrm{COR}(n)$ at time step $n = 10,000$ under one reward environment $\nu$: $(\mathrm{reward}^\dagger_\nu(10,000), \mathrm{COR}_\nu(10,000))$. The more the scatter plot is at the top of the graph, the higher the order estimation accuracy is, and the more the scatter plot is at the right, the greater the obtained reward is. \textbf{b} Time evolution of the average value of each metric over the whole ensemble of reward environments ($1\le n \le 10,000$).}
\label{fig:macro}
\end{figure}

\paragraph{Evaluation of the whole reward environments.}
Figure \ref{fig:macro}a summarizes the relationship between total rewards and order estimation accuracy: $x$-axis represents the normalized reward $\mathrm{reward}^\dagger(n)$, whereas $y$-axis represents the COR $\mathrm{COR}(n)$. Here, a normalized reward is defined as follows:
\begin{align*}
\mathrm{reward}^\dagger(n) := \frac{\mathrm{reward}(n)}{\mu^*\cdot n}
\end{align*}
Each plot in the graph indicates $\mathrm{reward}^\dagger(n)$ and $\mathrm{COR}(n)$ at time step $n = 10,000$ under one reward environment $\nu$: 
\[(\mathrm{reward}^\dagger_\nu(10,000), \mathrm{COR}_\nu(10,000))\]
Figure \ref{fig:macro}b shows the time evolution of the average value of each metric over the whole ensemble of reward environments from $n = 1$ to $n = 10,000$:
\begin{align*}
\frac{\sum_\nu\mathrm{COR}_\nu(n)}{\sum_\nu 1}, \quad
\frac{\sum_\nu\mathrm{reward}^\dagger_\nu(n)}{\sum_\nu 1}, \quad
(1\le n \le 10,000)
\end{align*}

\section*{Discussion}
\paragraph{Difficulty of maximizing rewards and arm order recognition.}
The results of the numerical simulations on the four-armed and eight-armed bandit problems show similar trends: there is a trade-off between the maximized total rewards and arm order recognition. As RoundRobin selects all arms equally, we always achieve a perfect COR at a time step $n = 10,000$ for any given reward environment. However, we cannot maximize rewards because regret linearly increases with time. On the contrary, in Chaos, we achieved normalized rewards of almost unity at the time step of $n = 10,000$ with respect to many types of reward environments. However, we can observe inferior performances regarding the arm order recognition accuracy because the arm selection is greatly biased to the best arm. In terms of the COR, the COR on RoundRobin and Chaos-CI (proposed method) quickly converged to unity. In terms of the total rewards, Chaos (previous method) and UCB1 are more active in using the exploitation principle to obtain greater rewards. The proposed method, Chaos-CI, achieves an outstanding performance on the arm order recognition and reward.

\begin{figure}
\centering
\includegraphics[width=0.7\linewidth]{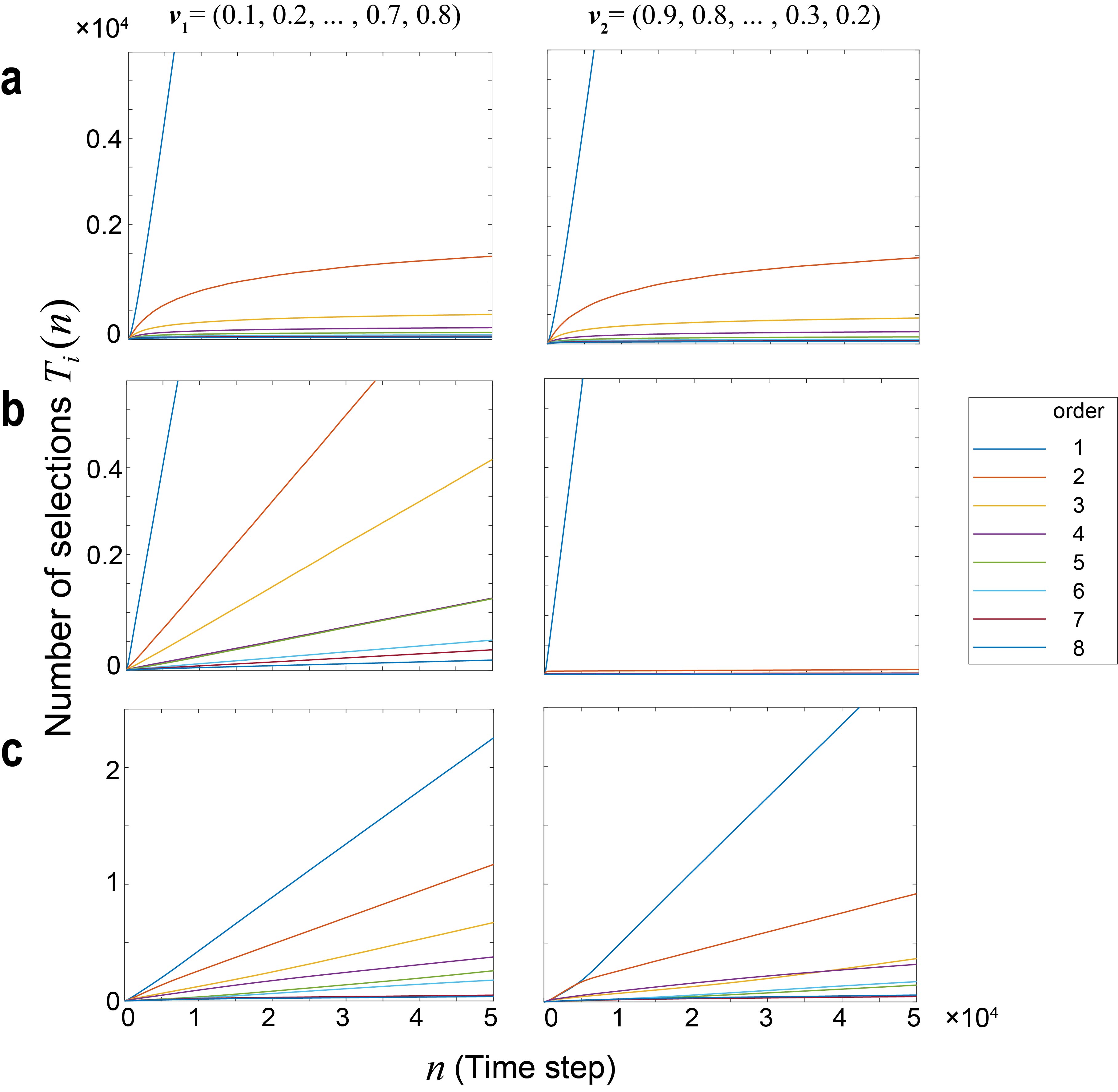}
\caption{Time evolution of the selection number of each arm for three methods: UCB1, Chaos, and Chaos-CI. The three figures on the left represent $T_i(n)$ under reward environment $\nu_1$, and the three on the right represent under $\nu_2$. Each row represents the result of each method: \textbf{a} UCB1, \textbf{b} Chaos, and \textbf{c} Chaos-CI.}
\label{fig:ti}
\end{figure}

\paragraph{Number of arm selections: $T_i(n)$.}
Figures \ref{fig:ti}a, b, and c show the time evolutions of $T_i(n)$ by UCB1, Chaos and Chaos-CI, respectively (RoundRobin leading to equal number of selections for all arms at any time). Here, we examine the two types of reward environments $\nu_1$ and $\nu_2$ in an eight-armed bandit given by $\nu_1:=(0.1,0.2,0.3,0.4,0.5,0.6,0.7,0.8)$ and $\nu_2:=(0.9,0.8,0.7,0.6,0.5,0.4,0.3,0.2)$ corresponding to the left and right columns of Fig. \ref{fig:ti}.

This figure shows that the selection number of the best arm (i.e. $T_{[1]}(n)$) increases by $O(n)$ and $T_i(n)$ $(i\neq i^*)$ increases almost by $O(\log n)$ in UCB1. Through the evolution of $T_i(n)$, UCB1 can achieve a regret of $O(\log n)$, but the convergence of ${\mathrm{COR}}(n)$ is slow. In the proposed Chaos-CI, the selection number of every arm evolves in a linear order. Therefore, the arm order recognition accuracy is faster than UCB1. Although the selections of non-top arms in the linear order cause regret to increase in a linear order, the slope of the linear-order regret is significantly decreased compared with that of RoundRobin by selecting better arms more often or by prioritizing the search (i.e. $T_{[1]}(n)>\cdots>T_{[K]}(n)$).

\paragraph{Environment dependency.}
As shown in Figs. \ref{fig:macro} and \ref{fig:ti}, the performances of Chaos are very different depending on reward environments $\nu_1$ and $\nu_2$. This finding is clearly linked with the arm selection number $T_i(n)$. In reward environment $\nu_1$, all $T_i(n)$ evolve in a linear order, but in reward environment $\nu_2$, $T_i(n)$ $(i\neq i^*)$ is approximately 100 at time step $n = 50,000$. Thus, the performance of Chaos heavily depends on the given reward environment. Table \ref{tab:result_var_met} summarizes the sample variance of metrics over 100 reward environments in an eight-armed bandit. As shown in the table, Chaos-CI is less dependent on reward environments and achieves more stable and higher accuracy than UCB1 and Chaos. In terms of obtained rewards, Chaos-CI has a larger variance than UCB1 and Chaos but is more stable than RoundRobin.

\begin{table}[htbp]
\begin{center}
\caption{Sample variance of metrics over 100 reward environments.($n = 10,000$)}
\label{tab:result_var_met}
\begin{tabular}{l|lc}
& COR & ${\rm reward}^\dagger$($\times10^{-3}$) \\ \hline
RoundRobin & 0 & 0.8852 \\
UCB1 & 0.0026 & 0.0116 \\
Chaos & 0.0413 & 0.3073 \\
Chaos-CI & 0.0004 & 0.6140
\end{tabular}
\end{center}
\end{table}

\section*{Conclusions}
In this study, we have examined ultrafast decision making with laser chaos time series in reinforcement learning (e.g. MAB) and set a goal to recognize the arm order of reward expectations by expanding the previous method, that is, time-division multiplexing of laser chaos recordings. In the proposed method, we have introduced exploration-degree adjustments based on confidence intervals of estimated rewards. The results of the numerical simulations based on experimental time series show that the selection number of each arm increases linearly, leading to a high and rapid order recognition accuracy.
Furthermore, arms with higher reward expectations are selected more frequently; hence, the slope of regret is reduced, although the selection number of an arm still linearly increases. Compared with UCB1 and Chaos, Chaos-CI (proposed method) is less dependent on the reward environment, indicating its potential significance in terms of robustness to environmental changes. In other words, Chaos-CI can make more accurate and stable estimates of arm order. Such an order recognition is useful in applications, such as channel selection and resource allocation in information and communications technology, where compromise actions or intelligent arbitrations are expected. 

\section*{Methods}
\subsection*{Optical system}
The device used was a distributed feedback semiconductor laser mounted on a butterfly package with optical fibre pigtails (NTT Electronics, KELD1C5GAAA). The injection current of the semiconductor laser was set to 58.5 mA (5.37$I_{th}$), where the lasing threshold $I_{th}$ was 10.9 mA. The relaxation oscillation frequency of the laser was 6.5 GHz, and its temperature was maintained at 294.83 K. The optical output power was 13.2 mW. The laser was connected to a variable fibred reflector through a fibre coupler, where a fraction of light was reflected back to the laser, generating high-frequency chaotic oscillations of optical intensity \cite{soriano2013relation,ohtsubo2012semiconductor,uchida2012optical}. The length of the fibre between the laser and reflector was 4.55 m, corresponding to a feedback delay time (round trip) of 43.8 ns. Polarization maintaining fibres were used for all of the optical fibre components. The optical signal was detected by a photodetector (New Focus, 1474-A, 38 GHz bandwidth) and sampled using a digital oscilloscope (Tektronics, DPO73304D, 33 GHz bandwidth, 100 GSample/s, eight-bit vertical resolution). The RF spectrum of the laser was measured by an RF spectrum analyzer (Agilent, N9010A-544, 44 GHz bandwidth). 

\subsection*{Details of the time-division multiplexing of laser chaos}
\begin{algorithm}[htbp]
\caption{Time-division multiplexing of laser-chaos \cite{naruse2018scalable}}
\label{code:chaos}
Parameters: $0<\alpha<1, \quad\Delta_S>0, \quad\Delta_L>0, \quad\Lambda>0, \quad\Omega>0$\\
Initialization: $\tau_S\leftarrow \tau_{init}, \quad\forall m,S_1,...,S_{m-1}: TH_{m,S_1\cdots S_{m-1}}\leftarrow 0$
\begin{algorithmic}[1]
\For{$n = 1,...,n_{\max}$}
\State $\tau_L\leftarrow \tau_S$
\State \% STEP1
\State $S_1\leftarrow \ifunc{s(\tau_L)\ge TH_1}$
\For{$m = 2,...,M$}
\State $\tau_L \leftarrow \tau_L+\Delta_L$
\State $S_m\leftarrow \ifunc{s(\tau_L)\ge TH_{m,S_1\cdots S_{m-1}}}$
\EndFor
\State \% STEP2
\State $a\leftarrow (S_1\cdots S_M)_2$
\State {\bf{sample}} $x_a(n) \sim P_a$
\State \% STEP3
\If{$x_{a}(n)>0$}
\For{$m = 1,...,M$}
\State $TH_{m,S_1\cdots S_{m-1}}\leftarrow \alpha TH_{m,S_1\cdots S_{m-1}} + \Lambda\cdot (-1)^{S_m}$
\EndFor
\Else
\For{$m = 1,...,M$}
\State $TH_{m,S_1\cdots S_{m-1}}\leftarrow \alpha TH_{m,S_1\cdots S_{m-1}} + \Omega\cdot (-1)^{1-S_m}$
\EndFor
\EndIf
\State $\tau_S \leftarrow \tau_S + \Delta_S$
\EndFor
\end{algorithmic}
\end{algorithm}

\paragraph{Convergence of Algorithm \ref{code:chaos}.}
For simplicity, we assume that $K = 2$ and the time series used for comparison with thresholds follows a uniform distribution of $[-1/2, 1/2]$ at an arbitrary time. We define the value of threshold $TH_1$ at the beginning of time step $n$ as $w(n)$. The time evolution of $w(n)$ can be represented as
\begin{align}
w(n+1) &= \alpha w(n) + q(n) \label{eq:w_rec}\\
&\text{where}\quad q(n) := \left\{\begin{array}{ll}
+ \Lambda & (\text{if $A(n)=0$, $X_0(n)=1$}) \\
- \Lambda & (\text{if $A(n)=1$, $X_1(n)=1$}) \\
+ \Omega & (\text{if $A(n)=1$, $X_1(n)=0$}) \\
- \Omega & (\text{if $A(n)=0$, $X_0(n)=0$}) 
\end{array}\right. \notag
\end{align}
The expectation of $w(n)$ is represented as follows.
\begin{align}
\Ex{q(n)} 
&= \Lambda \Prob{A(n)=0, X_0(n)=1} - \Lambda \Prob{A(n)=1, X_1(n)=1} \notag \\
&~+ \Omega \Prob{A(n)=1, X_1(n)=0} - \Omega \Prob{A(n)=0, X_0(n)=0} \notag \\
&= \Lambda\mu_0\Prob{A(n)=0}-\Lambda\mu_1\Prob{A(n)=1} +\Omega(1-\mu_1)\Prob{A(n)=1} - \Omega(1-\mu_0)\Prob{A(n)=0} \notag \\
&= \left\{(\Lambda + \Omega)\mu_0 - \Omega\right\}\Prob{A(n)=0} - \left\{(\Lambda + \Omega)\mu_1 - \Omega\right\}\Prob{A(n)=1} \label{eq:qn_ex}
\end{align}
Because we assume that $s(t)$ follows a uniform distribution, if $\max\{n\Lambda,n\Omega\}<1/2$, 
\begin{align*}
\Prob{A(n)=0} 
&= \integ{}{}{\Prob{A(n)=0|w(n)=x}\cdot\Prob{w(n)=x}}{x} \\
&= \integ{}{}{\left(\frac{1}{2} + x\right)\Prob{w(n)=x}}{x} \\
&= \frac{1}{2} + \Ex{w(n)} \\
\Prob{A(n)=1}
&= \integ{}{}{\Prob{A(n)=1|w(n)=x}\cdot\Prob{w(n)=x}}{x} \\
&= \integ{}{}{\left(\frac{1}{2} - x\right)\Prob{w(n)=x}}{x} \\
&= \frac{1}{2} - \Ex{w(n)}
\end{align*}
By Eqs. (\ref{eq:w_rec}) and (\ref{eq:qn_ex}),
\begin{align}
\Ex{w(n+1)} 
&= \alpha\Ex{w(n)} + \Ex{q(n)} \notag \\
&= \frac{1}{2}(\Lambda + \Omega)(\mu_0 - \mu_1) + \left\{\alpha + (\Lambda + \Omega)(\mu_0 + \mu_1)-2\Omega\right\}\Ex{w(n)} \label{eq:w_ex_rec}
\end{align}
Equation (\ref{eq:w_ex_rec}) can lead to
\begin{align}
\Ex{w(n)} = \frac{P}{1-Q} + Q^{n-1}\left(\Ex{w(1)}-\frac{P}{1-Q}\right) \label{eq:wn_ex}
\end{align}
where
\begin{align*}
P := \frac{1}{2}(\Lambda + \Omega)(\mu_0 - \mu_1),\quad
Q := \alpha + (\Lambda + \Omega)(\mu_0 + \mu_1)-2\Omega
\end{align*}
Equation (\ref{eq:wn_ex}) indicates that $\Prob{A(n)=0}$ and $\Prob{A(n)=1}$ converge to a certain value in $(0,1)$ if $|P/(1-Q)|<1/2$ and $|Q|<1$. In this case, the number of selections for each arm linearly increases. Furthermore, if $|P/(1-Q)|\ge1/2$ or $|Q|\ge 1$, then convergence or divergence occurs at $|\Ex{w(n)}|\ge1/2$, which leads to $\Prob{A(n)=1}\approx 1$ or $\Prob{A(n)=0}\approx 1$. In this case, one of the arms will be selected intensively as time passes.

The above discussion shows that the convergence and performance of Algorithm \ref{code:chaos} depend on learning rate $\alpha$, exploration degree $(\Lambda,\Omega)$, and reward environment $(\mu_0,\mu_1)$.

\subsection*{Details of the proposed method}
\begin{algorithm}[htbp]
\caption{Proposed method}
\label{code:chaos_ci}
Parameters: $0<\alpha<1, \quad\Delta_S>0, \quad\Delta_L>0, \quad d\in\mathbb{N}, \quad \beta > 1$\\
Initialization: \\
\quad $\tau_S\leftarrow \tau_{init}, \quad t_i \leftarrow 0, \quad r_i \leftarrow 0$, \\
\quad $\forall m,S_1,...,S_{m-1}: TH_{m,S_1\cdots S_{m-1}}\leftarrow 0, \quad \Lambda_{m,S_1\cdots S_{m-1}}\leftarrow \Lambda_{init}, \quad \Omega_{m,S_1\cdots S_{m-1}}\leftarrow \Omega_{init}$
\begin{algorithmic}[1]
\For{$n=1,...,n_{\max}$}
\State $\tau_L\leftarrow \tau_S$
\State $\vdots$
\State (Algorithm\ref{code:chaos}: STEP1 -- STEP3)
\State $\vdots$
\State $t_a \leftarrow t_a + 1, \quad r_a \leftarrow r_a + x_a$
\If{$n\mod d=0$}
\For{$m=1,...,M$}
\If{two confidence intervals are overlapped,}
\State $\Lambda_{m,S_1\cdots S_{m-1}} \leftarrow \Lambda_{m,S_1\cdots S_{m-1}} / \beta$
\State $\Omega_{m,S_1\cdots S_{m-1}} \leftarrow \Omega_{m,S_1\cdots S_{m-1}}/\beta $
\Else
\State $\Lambda_{m,S_1\cdots S_{m-1}} \leftarrow \Lambda_{m,S_1\cdots S_{m-1}} \cdot \beta$
\State $\Omega_{m,S_1\cdots S_{m-1}} \leftarrow \Omega_{m,S_1\cdots S_{m-1}}\cdot\beta $
\EndIf
\EndFor
\EndIf
\State $\tau_S \leftarrow \tau_S + \Delta_S$
\EndFor
\end{algorithmic}
\end{algorithm}

\paragraph{Convergence of the proposed method.}
In the previous paragraph, we have found that the performance of the algorithm proposed is heavily dependent on parameters $(\Lambda, \Omega)$. Therefore, in the proposed method, exploration-degree adjustments based on confidence intervals are added to Algorithm \ref{code:chaos}: if the exploration itself is not sufficient, then thresholds are set close to 0 and values of $(\Lambda, \Omega)$ decrease, so thresholds are less likely to diverge, which leads to improved accuracy. If exploration is applied sufficiently, then the values of $(\Lambda, \Omega)$ increase, so the thresholds are more likely to diverge, which leads to an intensive selection of a better arm and slow increase of regret.

\section*{Data availability}
The datasets generated during the current study are available from the corresponding author on reasonable request.

\bibliography{ref}

\section*{Acknowledgements}

This work was supported in part by the CREST project (JPMJCR17N2) funded by the Japan Science and Technology Agency and Grants-in-Aid for Scientific Research (A) (JP17H01277) funded by the Japan Society for the Promotion of Science. The authors acknowledge Atsushi Uchida and Kazutaka Kanno for the measurements of laser chaos time series and Satoshi Sunada and Hirokazu Hori for their variable discussions about chaos and order recognition.

\section*{Author contributions}

M.N. directed the project and N.N. designed the order recognition algorithm and conducted signal processing. N.N., N.C., M.H., and M.N. analyzed the data and N.N. and M.N. wrote the paper. 

\section*{Competing interests }
The authors declare no competing interests.

\section*{Additional information }
Correspondence and requests for materials should be addressed to N.N. and M.N.

\end{document}